\documentclass[%
 reprint,
 superscriptaddress,
 preprintnumbers,
 amsmath,amssymb,
 aip,
]{revtex4-1}

\usepackage[utf8]{inputenc}
\usepackage{graphicx}
\usepackage{grffile}
\usepackage[usenames,dvipsnames]{xcolor}
\usepackage{amsmath}
\usepackage[normalem]{ulem}
\usepackage[resetlabels, labeled]{multibib}
\usepackage{appendix}
\newcites{S}{References Supplementary Materials}
\definecolor{orange}{rgb}{1,0.5,0}
\definecolor{goodgreen}{rgb}{0.1,0.5,0}
\definecolor{goodred}{rgb}{0.7,0,0}
\usepackage{lineno}
\usepackage{xcolor}
\setpagewiselinenumbers
\modulolinenumbers[5]

\usepackage[colorlinks,urlcolor=goodgreen,citecolor=blue,linkcolor=goodred]{hyperref}
\usepackage{float}
\usepackage{babel}
\graphicspath{ {images/} }

\makeatother

\begin{document}


\title[JTWPA with a 2$^\text{nd}$ harmonic CPR]{Effect of a 2$^\text{nd}$ harmonic current--phase relation on the behavior of a Josephson Traveling Wave Parametric Amplifier}

\newcommand{\orcid}[1]{\href{https://orcid.org/#1}{\includegraphics[width=8pt]{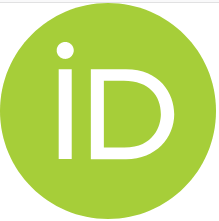}}}
\author{Claudio Guarcello\orcid{0000-0002-3683-2509}}
\email{Author to whom correspondence should be addressed: cguarcello@unisa.it}
\affiliation{Dipartimento di Fisica ``E.R. Caianiello'', Universit\`a di Salerno, Via Giovanni Paolo II, 132, I-84084 Fisciano (SA), Italy}
\affiliation{INFN, Gruppo Collegato di Salerno, Via Giovanni Paolo II, 132, I-84084 Fisciano (SA), Italy}
\author{Carlo Barone\orcid{0000-0002-6556-7556}}
\email{cbarone@unisa.it }
\affiliation{Dipartimento di Fisica ``E.R. Caianiello'', Universit\`a di Salerno, Via Giovanni Paolo II, 132, I-84084 Fisciano (SA), Italy}
\affiliation{INFN, Gruppo Collegato di Salerno, Via Giovanni Paolo II, 132, I-84084 Fisciano (SA), Italy}
\author{Giovanni Carapella\orcid{0000-0002-0095-1434}}
\email{gcarapella@unisa.it }
\affiliation{Dipartimento di Fisica ``E.R. Caianiello'', Universit\`a di Salerno, Via Giovanni Paolo II, 132, I-84084 Fisciano (SA), Italy}
\affiliation{INFN, Gruppo Collegato di Salerno, Via Giovanni Paolo II, 132, I-84084 Fisciano (SA), Italy}
\author{Giovanni Filatrella\orcid{0000-0003-3546-8618}}
\email{filatr@unisannio.it}
\affiliation{Science and Technology Department, University of Sannio, Benevento, Italy}%
\author{Andrea Giachero\orcid{0000-0003-0493-695X}}
\email{andrea.giachero@unimib.it }
\affiliation{Department of Physics, University of Milano Bicocca, Piazza
della Scienza, I-20126 Milano, Italy}
\affiliation{INFN—Milano Bicocca, Piazza della Scienza, I-20126 Milano, Italy}
\affiliation{Bicocca Quantum Technologies (BiQuTe) Centre, Piazza della Scienza, I-20126 Milano, Italy}
\author{Sergio Pagano\orcid{0000-0001-6894-791X}}
\email{spagano@unisa.it }
\affiliation{Dipartimento di Fisica ``E.R. Caianiello'', Universit\`a di Salerno, Via Giovanni Paolo II, 132, I-84084 Fisciano (SA), Italy}
\affiliation{INFN, Gruppo Collegato di Salerno, Via Giovanni Paolo II, 132, I-84084 Fisciano (SA), Italy}

\date{\today}

\begin{abstract}
We numerically investigate the behavior of a Josephson traveling wave parametric amplifier assuming a current-phase relation with a second--harmonic contribution. We find that varying the weight of harmonic terms in the Josephson current affects the gain profile. The analysis of gain characteristics, phase-space portraits, Poincar\'e sections, and Fourier spectra demonstrates that the nonsinusoidal contribution influences the operating mode and stability of the device. In particular, we identify the optimal weighting of harmonic contributions that maximizes amplification, achieving gains up to $\sim 13\;\text{dB}$ in a device without dispersion engineering.
\end{abstract}

\maketitle

Boosting weak microwave signals is essential in many application fields, like reading superconducting qubits, quantum devices, and radio astronomy~\cite{Bartram2023,Shiri2024,Citro2024}. Optimal low-noise microwave amplifiers get close to the quantum--noise limit by working at very low temperatures and using parametric pumping in circuits with Josephson junctions (JJs) or superconducting elements with high kinetic inductance~\cite{Aumentado2020}.

Superconducting amplifiers based on nonlinear resonators usually have reasonable gain and quantum-limited noise, but feature a limited bandwidth of a few hundred MHz. Alternatively, the so-called \emph{traveling-wave parametric amplifiers} (TWPAs) can achieve much wider bandwidths, up to several $\text{GHz}$~\cite{Cullen1960,Sorenssen1962,Esposito2021}. This aspect is especially useful for reading multiple qubits and detection applications~\cite{Pagano2022}. The efficient use of quantum hardware through frequency multiplexing is a crucial factor for the scalability of quantum processors. However, superconducting TWPAs have also drawbacks, such as gain ripples, generation of unwanted tones and comparatively lower gain. These challenges need to be addressed by proper design of the distributed amplifiers and control of the intrinsic nonlinearities.

Superconducting TWPAs are currently realized using the high nonlinear kinetic inductance found in superconductors~\cite{Bockstiegel2014,Vissers2016,Chaudhuri2017,Ranzani2018,Malnou2021} or in the JJs embedded in the transmission line. In the latter case they are called Josephson TWPAs (JTWPAs)~\cite{Sweeny1985,Bell2015,Zorin2016,Fasolo2019,Esposito2021,Esposito2022}. One of the primary goals in designing a JTWPA is determining strategies that maximize signal amplification, while keeping a good control of the system bandwidth and of spurious tones generation. The amplifier's \emph{gain} can be optimized through dispersion engineering, e.g., by \emph{resonant phase matching} strategies~\cite{O'Brien2014,White2015,Macklin20215} involving periodically embedded LC resonators capacitively coupled to the transmission line: the resonators adapt the dispersion relation so that the total phase mismatch along the transmission line remains small over a wide bandwidth. 
Other dispersion--engineering strategies to maximize the achievable gain and the robustness of the device response can be also implemented~\cite{Winkel2020,Planat2020,Gaydamachenko2022,Roudsari2023,Qiu2023,Rizvanov2024}. 
The basic element of a JTWPA, i.e., the Josephson junction, determines the device performance. Usually, a JJ is assumed to follow a sinusoidal current-phase relation (CPR). However, this is not always the case, as higher harmonics are often observed in the CPR of unconventional JJs~\cite{Carbone, Thompson}. 
For instance, in nanobridge-based junctions, deviations from a sinusoidal CPR at high temperatures to a sawtooth-like or even multivalued form at lower temperatures arise due to phase slippage effects in the weak link~\cite{Troeman2008}.
In this work, we precisely focus on the gain profiles in the specific case of a CPR with a second--harmonic contribution. 
\begin{figure*}[t!!]
\includegraphics[width=2\columnwidth]{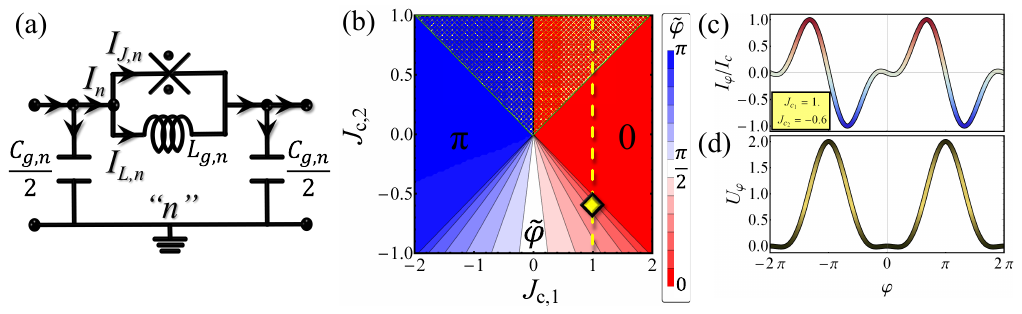}
\caption{(a) $n$-th cell of the JTWPA. (b) GS map, $\widetilde{\varphi}\left(J_{c_1},J_{c_2}\right)$. (c) CPR, $I (\varphi)$, and (d) Josephson energy, $U (\varphi)$, at $J_{c_1}=1$ and $J_{c_2}=-0.6$.}\label{Fig01}
\end{figure*}

To do so, we numerically investigate the behavior of a JTWPA~\cite{GuarcelloJTWPA2023,GuarcelloJTWPA2024} with the realistic system parameter setting from the design developed within the DARTWARS INFN collaboration~\cite{Giachero2022,Pagano2022,Granata2023,Borghesi2023,Rettaroli2023,Fasolo2024,Ahrens2024,Faverzani2024}. In particular, we consider a discrete transmission line formed by 990 cells, each formed by a capacitor $C_{g,n} = 24\, \text{fF}$ to ground and an rf--SQUID in series along the signal path, see Fig.~\ref{Fig01}(a). Each rf--SQUID consists of one JJ in parallel with an inductor $L_{g,n} = 120\, \text{pH}$. At the input of the transmission line is connected a voltage source, having a standard internal impedance, $R_i = 50\, \Omega$, and generating pump and signal voltages, $V_{\text{pump}}$ and $V_{\text{sign}}$, respectively.
The end of the transmission line is connected, through a series capacitance, $C_{\ell} = 1\, \text{nF}$, for DC decoupling, a standard load impedance $R_{\ell} = 50\, \Omega$, across which the voltage drop, $V_{\text{out}}$, is measured. Setting of boundary conditions is highly relevant when studying Josephson transmission lines, for load--matched boundary conditions can significantly impact the dynamics and even suppress chaos~\cite{Pankratov2008,Pankratov2017}.
The current through the JJ of the $n$-th cell is expressed by the well-known resistively and capacitively shunted junction (RCSJ) model~\cite{Stewart1968,McCumber1968,Barone1982,Piedjou2021,Guarcello2021_PRAppl,Grimaudo2022} as
\begin{equation}\label{RCSJ}
I_{J,n} = C_J \frac{\hbar}{2e} \frac{d^2 \varphi_n}{dt^2} + \frac{1}{R_J} \frac{\hbar}{2e} \frac{d\varphi_n}{dt} + I(\varphi_n),
\end{equation}
where we assume $C_J = 200\, \text{fF}$ and $R_J = 20\, k\Omega$. 
We deal with the numerical solution of the system of coupled differential equations, one for each cell of the JTWPA, with appropriate boundary conditions, through an implicit finite--difference method based on a tridiagonal algorithm, i.e., a common route for numerically studying Josephson transmission lines~\cite{Lomdahl1982,Fedorov2008,Guarcello2018,Guarcello2021,Guarcello2024_APL,Guarcello2024_CSF}. The time step and the integration period, in units of the Josephson plasma frequency~\cite{Barone1982}, are $\Delta t = 10^{-2}$ and $t_{\text{max}} = 2\times10^4$, respectively. Numerical details can be found in Ref.~\onlinecite{Guarcello2024}. 

We are interested in exploring the JTWPA performance in the case of a CPR with a second--harmonic contribution, i.e.,
\begin{equation}\label{I(varphi)}
I (\varphi) = J_{c_1} \sin(\varphi) + J_{c_2} \sin(2\varphi).
\end{equation}
This type of CPR has demonstrated to well describe superconducting junctions with ferromagnetic barriers~\cite{Barash1995,Tanaka1996,Buzdin2003,Goldobin2007,Goldobin2011,Pal2014,Goldobin2015,Sickinger2012,Menditto2016,Stoutimore2018}, as well as JJs based on semiconductors~\cite{Baumgartner2022,Pal2022,Sivakumar2024,Zhang2024,Leblanc2024,Ciaccia2024,Valentini2024,Leblanc2024b,Zhang2025}, graphene~\cite{Messelot2024}, or unconventional superconductors~\cite{Il'ichev1999a,Il'ichev1999b}. Higher-order contributions, typically overlooked in conventional tunnel JJs, have recently been demonstrated to affect also the excited--state spectrum of Al-- and AlO--based Josephson qubits~\cite{Willsch2024}. Finally,  a recent proposal suggests that also altermagnets could provide an alternative route for CPR engineering~\cite{Sun2024}.

The Josephson energy associated to the CPR in Eq.~\eqref{I(varphi)} reads 
\begin{equation}\label{U(varphi)}
U(\varphi) = -J_{c_1} \cos(\varphi) - J_{c_2} \cos(2\varphi)/2.
\end{equation}
The phase, $\widetilde{\varphi}$, of the ground state (GS) is determined by minimizing $U(\varphi)$, which requires solving the equation
\begin{equation}
\left . \frac{\partial U}{\partial \varphi} \right |_{\varphi=\widetilde{\varphi}} = \sin(\widetilde{\varphi}) \left [ J_{c_1} + 2 J_{c_2} \cos(\widetilde{\varphi}) \right ] = 0,
\end{equation}
so that the values of $\widetilde{\varphi}$ are given by
\begin{equation}
\sin(\widetilde{\varphi}) = 0 \qquad \text{or} \qquad \cos(\widetilde{\varphi}) = -\frac{1}{2g}.
\end{equation}
Here, $g = J_{c_2}/J_{c_1}$ is the ratio of the weights of the harmonic contributions.
Therefore, there are also non--trivial GSs, i.e., $\widetilde{\varphi} = 0$, $\pi$, or $\widetilde{\varphi}\in(0-\pi)$. In detail, when $\left| g\right|<1/2$, the GS can be only $\widetilde{\varphi} = 0$, if $J_{c_1} > 0$, or $\widetilde{\varphi} = \pi$, if $J_{c_1} <0$. Instead, if $\left| g\right|\geq 1/2$, the system has non--trivial GSs, $\widetilde{\varphi}=\pm\arccos\left(1/2g\right)$, if $J_{c_2} < 0$.

\begin{figure*}[t!!]
\includegraphics[width=2\columnwidth]{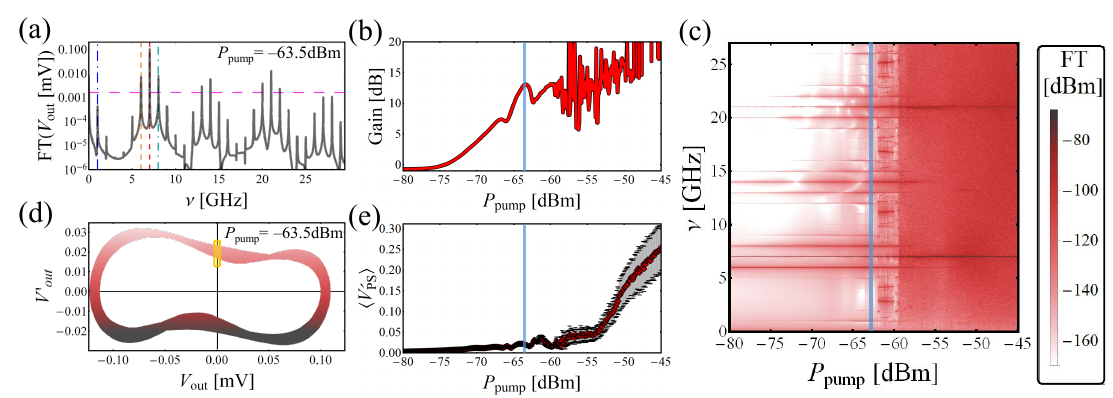}
\caption{(a) FT of $V_{\text{out}}$ at $P_{\text{pump}}= -63.5\;\text{dBm}$. (b) Gain versus $P_{\text{pump}}$. (c) Fourier map. (d) Phase portrait, $V_{\text{out}}'$ \emph{versus} $V_{\text{out}}$, at $P_{\text{pump}}= -63.5\;\text{dBm}$. (e) Poincar\'e section, $\left<V'_{\text{PS}} \right>$, versus $P_{\text{pump}}$. The other parameters are: $\nu_{\text{pump}} = 7\;\text{GHz}$, $\nu_{\text{sign}}=6\;\text{GHz}$, $P_{\text{sign}} = -100\;\text{dBm}$, and $I_{\text{bias}}=0$. In (a), the red (orange) dashed vertical line marks $\nu_{\text{pump}}$ ($\nu_{\text{sign}}$), the cyan (blue) dot-dashed vertical line indicates the 4WM (3WM) $\nu_{\text{idle}}$, and the magenta horizontal line indicates the input--signal level. In (d), the region in yellow highlights the distribution of $V_{\text{out}}'$ collected to construct the plot in (e). In (b), (c) and (e), a vertical, cyan line indicates the value $P_{\text{pump}}= -63.5\;\text{dBm}$ set in (a) and (d) (Multimedia available online).}\label{Fig02}
\end{figure*}

Figure~\ref{Fig01}(b)~\cite{Goldobin2007} shows the GS map, $\widetilde{\varphi}\left(J_{c_1},J_{c_2}\right)$. In the following, we explore the JTWPA response along the vertical, yellow dashed line in this plot, i.e., for $J_{c_1}=1$ and $J_{c_2}\in[-1,1]$.
The region for $J_{c_2} > 0$ and $\left| g\right|\geq1/2$ is highlighted since it corresponds to $U(\varphi)$ profiles with an additional secondary minimum in $\varphi=0$ or $\pi$~\cite{Goldobin2007}. Figures~\ref{Fig01}(c) and (d) present, respectively, the CPR and the Josephson energy at $\left(J_{c_1},J_{c_2}\right)= \left(1,-0.6\right)$, i.e., the point marked by a yellow diamond in the GS map in Fig.~\ref{Fig01}(a). It is evident that the minimum energy is slightly off--center with respect to $\varphi=0$. 

The physical meaning of negative $J_{c_1}$ and $J_{c_2}$ is intertwined with the mechanisms governing supercurrent transport. A negative $J_{c_1}$ typically indicates a $\pi$-junction~\cite{Birge2024}, commonly found in systems such as superconductor-ferromagnet-superconductor (SFS) junctions, where the exchange interaction in the ferromagnetic layer induces oscillations in the superconducting wavefunction~\cite{Golubov2004,Ryazanov2001}. In unconventional superconductors, such as $d$-wave materials, a similar effect, depending on the junction's orientation relative to the superconducting gap symmetry, occurs. In high-$T_c$ cuprates, grain-boundary JJs exhibit this phenomenon, leading to spontaneous half-flux quantum states in superconducting loops~\cite{VanHarlingen1995,Kirtley1999}. 
A negative $J_{c_2}$ suggests a more intricate interplay of transport channels and becomes particularly relevant when $J_{c_1}$ is suppressed, such as near the 0–$\pi$ transition in SFS junctions~\cite{Kontos2001}. Ferromagnetic junctions thus provide a direct way to intrinsically ``control'' $g$ by adjusting ferromagnetic layer properties~\cite{Sickinger2012,Menditto2016}. 
A recently introduced family of logics exploits JJs with a dominant 2nd-harmonic component in their CPR~\cite{Soloviev2021,Jabbari2023}. In the original proposal, phase logic circuits were based on bistable junctions with $I_s(\varphi) \propto \sin(2\varphi)$, offering a promising route for high--density superconducting circuits. Notably, a 0–0–$\pi$ SQUID, i.e., the building block of the so-called \emph{half-flux-quantum logic}, is effectively equivalent to a single junction with a CPR containing a negative 2nd-harmonic term, $I_s(\varphi) \propto -  \sin(2\varphi)$~\cite{Soloviev2021,Takeshita2023}.

The maximum supercurrent, i.e., the \emph{critical current}, $ I_c=\max_{\varphi} \left\{I(\varphi) \right\}$, depends on $g$. In units of $J_{c_1}$, Eq.~\eqref{I(varphi)} reads
\begin{equation}\label{gammavarphi}
\gamma(\varphi,g) = \sin(\varphi) + g \sin(2\varphi),
\end{equation}
with maxima, $\gamma_\pm(g)$, given by~\cite{Goldobin2007}
\begin{equation}
\gamma_{\pm}(g) \!=\! \frac{1}{32|g|} \!\left( \sqrt{1 + 32g^2 }\pm 3 \right)^{\!\frac{3}{2}}\!\!\left( \sqrt{1 + 32g^2} \mp 1 \right)^{\!\frac{1}{2}}\!.
\end{equation}
It turns out that $\gamma_{+}(g) > \gamma_{-}(g)$ (in particular, $\Delta \gamma = \gamma_{+} - \gamma_{-} \approx \sqrt{2}$), with the secondary maximum, $\gamma_{-}(g)$, appearing only if $|g| \geq 1/2$.
Therefore, the critical current is $I_c = J_{c_1} \gamma_{+}$, so that the Josephson current can be recast as $I(\varphi) = I_c \times i(\varphi, g)$. Here, we stress the role of the normalized current $i(\varphi, g) = \gamma(\varphi,g)/\gamma_{+}(g)\in[-1, +1]$, which embodies the $\varphi$-dependence.
Indeed, we aim to scan the $J_{c_2}$ values along the path marked by the yellow line in Fig.~\ref{Fig01}(a), i.e., at $J_{c_1}=1$ so that $J_{c_2}\equiv g$~\cite{Note1}
. However, modifying $g$ also changes the critical current, which may significantly influence the performance of the JTWPA (for example, changing $I_c$ affects both the Josephson plasma frequency and the Josephson inductance, e.g., eventually leading to an unwanted load mismatch). Instead, we want to show how the shape of the CPR, rather than the critical current itself, can lead to a non--trivial response in the JTWPA. Therefore, in what follows, we fix $I_c = 2\,\mu\text{A}$, a value consistent with the device specifications reported in Ref.~\onlinecite{Pagano2022}, and we investigate how the parameter $g$ affects the JTWPA performance assuming different normalized CPRs $i(\varphi, g)$.

\begin{figure*}[t!!]
\includegraphics[width=1.5\columnwidth]{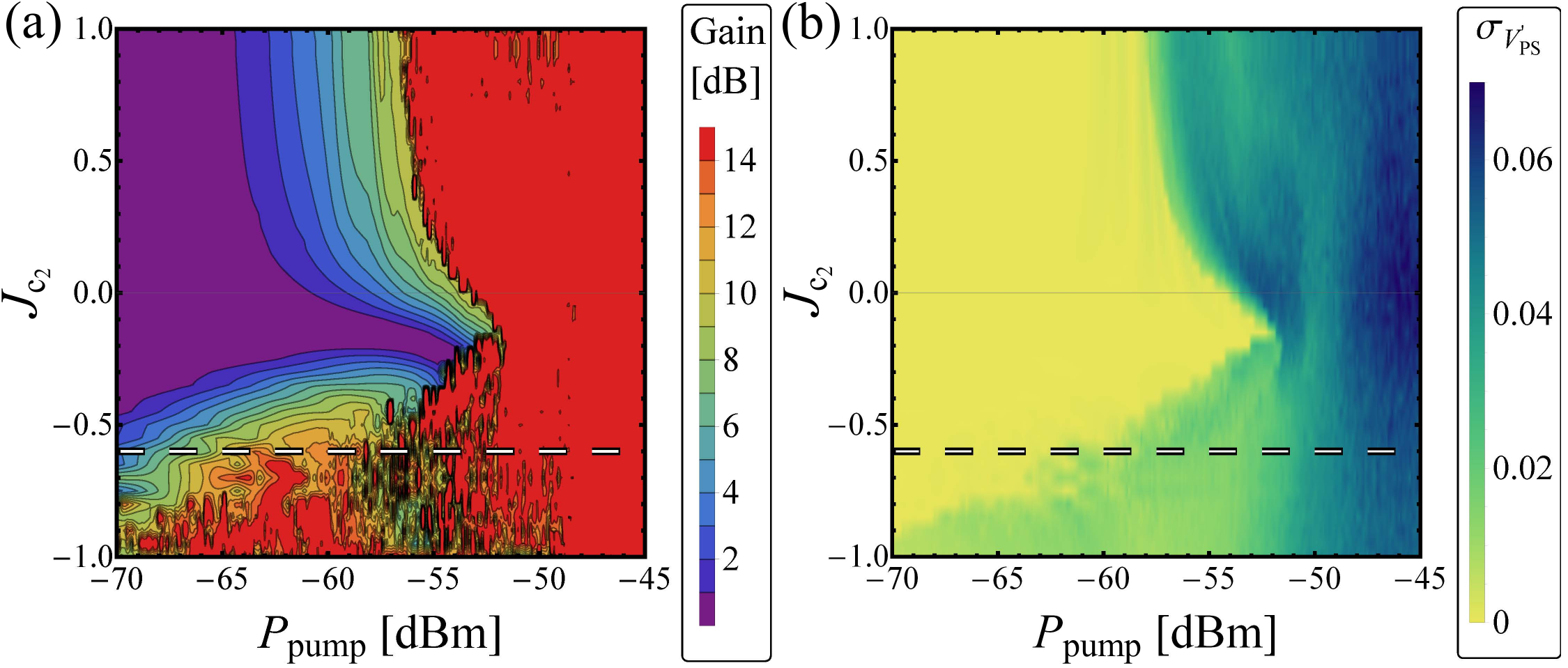}
\caption{(a) Gain and (b) $V'_{\text{PS}}$--standard deviation, $\sigma_{V'_{\text{PS}}}$, as a function of $J_{c_2} \in [-1, 1]$ and $P_{\text{pump}}\in[-70,-45]\;\text{dBm}$. The other parameters are: $\nu_{\text{pump}} = 7\;\text{GHz}$, $\nu_{\text{sign}}=6\;\text{GHz}$, $P_{\text{sign}} = -100\;\text{dBm}$, and $I_{\text{bias}}=0$. The white horizontal dashed line marks the value $J_{c_2} =-0.6$ used for Fig.~\ref{Fig02}.}\label{Fig03}
\end{figure*}

We first look at the Fourier transform of the output signal, from which we extract the gain in dB, defined as $\text{Gain} = 20 \log [V_{\text{out}}(\nu_{\text{sign}})/V_{\text{sign}}]$. Figure~\ref{Fig02} shows the results for $J_{c_1}= 1$ and $J_{c_2}= -0.6$ (Multimedia available online). We will demonstrate in the following that this value of $J_{c_2}$ maximizes the gain with stable conditions. We consider the effects produced by the change in pump intensity, $P_{\text{pump}}$, taking the other driving parameters fixed, that is, $\nu_{\text{pump}} = 7\;\text{GHz}$, $\nu_{\text{sign}}=6\;\text{GHz}$, $P_{\text{sign}} = -100\;\text{dBm}$, and $I_{\text{bias}}=0$. In Figs.~\ref{Fig02}(b), (c), and (e), a vertical, cyan line indicates the value $P_{\text{pump}}= -63.5\;\text{dBm}$ set in (a) and (d).

In Fig.~\ref{Fig02}(a) we show the \emph{Fourier transform} (FT) of $V_{\text{out}}$ at $P_{\text{pump}}= -63.5\;\text{dBm}$; the red (orange) dashed vertical line marks the pump (signal) frequency, while the cyan dot--dashed line indicates the \emph{idle tone} at frequency $\nu_{\text{idle}}=2 \nu_{\text{pump}} - \nu_{\text{sign}} =8\;\text{GHz}$ in the so--called four--wave mixing (4WM) regime. Interestingly, even the three--wave mixing (3WM) mode at $\nu_{\text{idle}}=\nu_{\text{pump}} - \nu_{\text{sign}}=1\;\text{GHz}$ emerges clearly (see the blue dot-dashed vertical line), despite the absence of bias current and magnetic field. The horizontal magenta dashed line marks the input--signal level: the amplification affecting the signal tone is quite evident. Thus, we compute the signal gain by ranging the pump intensity in $P_{\text{pump}}\in [-80,-45]\;\text{dBm}$, see Fig.~\ref{Fig02}(b). 
We can identify three distinct regimes: \emph{i}) for $P_{\text{pump}} \lesssim -62.5\;\text{dBm}$, the gain increases (non--monotonically), reaching values above $\sim 13\;\text{dB}$, then \emph{ii}) for $P_{\text{pump}}\in(-62.5, -59.5)\;\text{dBm}$, the gain smoothly rises again, and \emph{iii}) for $P_{\text{pump}} \gtrsim -59.5\;\text{dBm}$, the gain profile becomes highly scattered, indicating the onset of a chaotic response~\cite{Guarcello2024}.
This distinction is particularly clear when looking at \emph{Fourier map} in Fig.~\ref{Fig02}(c), i.e., the frequency components in the output signal, with peaks corresponding to the dominant frequencies. At low pump powers, the spectrum is relatively sparse, although several harmonics can be observed. As $P_{\text{pump}}$ increases, additional spectral components appear, and the harmonic peaks become more pronounced, indicating non-linear effects such as harmonic generation. For $P_{\text{pump}} \gtrsim -59.5\;\text{dBm}$, the spectrum broadens significantly, implying a higher noise level and a transition towards chaotic behavior in the system's dynamics~\cite{Guarcello2024}.

Figure~\ref{Fig02}(d) shows the \emph{phase-space portraits}, $V_{\text{out}}'$ versus $V_{\text{out}}$, at $P_{\text{pump}}= -63.5\;\text{dBm}$. It is characterized by a relatively broad trajectory, an aspect tightly intertwined with the FT response, which displays multiple peaks. From this plot, we can construct the Poincar\'e section by collecting, for each parameter combination, all values of $V_{\text{out}}'$ when $V_{\text{out}}=0$, i.e., the trajectory crossings within the region highlighted in yellow in Fig.~\ref{Fig02}(d).
In this way, for each pump intensity, we obtain a ``distribution" of \( V_{\text{out}}' \) values, represented as points in the Poincaré section. When these points form tightly concentrated clusters, this may indicate the presence of narrow and stable periodic orbits in the phase portrait. Conversely, when the points are widely dispersed, it suggests transitions to chaotic behavior, which is reflected in a tangled, intricate structure within the phase portrait.
From each distribution at a given $P_{\text{pump}}$, we extract the mean and the standard deviation, which are then used to construct the plot in Fig.~\ref{Fig02}(e), showing $\left<V'_{\text{PS}} \right>$ as a function of $P_{\text{pump}}$. The error bars represent the $V'_{\text{PS}}$--standard deviation, $\sigma_{V'_{\text{PS}}}$. Once again, the distinct regimes, each exhibiting markedly different responses, are clearly visible.

To complement and enrich the study of the JTWPA's performance, the multimedia file available online presents a collection of results, similar to those in Fig.~\ref{Fig02}. This figure is essentially a still frame at $J_{c_2} = -0.6$ from the animation, which is instead obtained by sweeping $J_{c_2} \in [-1, 1]$ with steps of $\Delta J_{c_2} = 0.05$, i.e., along the yellow dashed line in Fig.~\ref{Fig01}(a).

To better understand the amplification mechanism as $J_{c_2}$ changes, Fig.~\ref{Fig03} shows (a) the gain and (b) the $V'_{\text{PS}}-$standard deviation, $\sigma_{V'_{\text{PS}}}$, as a function of $J_{c_2} \in [-1, 1]$ and $P_{\text{pump}} \in [-70, -45]\;\text{dBm}$. Specifically, in Fig.~\ref{Fig03}(a) we show the $\text{Gain}(P_{\text{pump}},J_{c_2})$ map: a red region marks parameter combinations that lead to a chaotic response. The value of $J_{c_2}$ has a significant impact on both the maximum achievable gain and the range of $P_{\text{pump}}$ values within which the system exhibits a non-chaotic behavior. In fact, for $J_{c_2}>0$, the gain shows little variation as $J_{c_2}$ changes. 
In contrast, a markedly different response emerges for $J_{c_2}<0$. 
First, we observe that the largest range of stable $P_{\text{pump}}$ values occurs for $J_{c_2} \simeq -0.15$. Then, as $J_{c_2}$ decreases further, the gain reaches higher values, in particular close to $J_{c_2} = -0.6$ (the horizontal, white dashed line), but the system becomes more prone to chaotic oscillations, as highlighted by the jagged transition between stable and unstable regions. This instability manifests as a narrower range of $P_{\text{pump}}$ values that allow for a stable response as $J_{c_2}$ decreases. The system's sensitivity to changes in $J_{c_2}$ is clearly visible in this region of the $(P_{\text{pump}},J_{c_2})$-parameter space, revealing the delicate balance between high gain and stability, especially with a non--trivial GS. This threshold value emerges also by expanding $\gamma(\varphi, g)$ for small $\varphi$, yielding $\gamma(\varphi, g) \simeq (1+2g)\varphi - (1+8g)\varphi^3$. At $g = -1/2$ and $g = -1/8$, the coefficients of the first- and third-order terms change sign, respectively. It is therefore reasonable to expect that these points mark sharp transitions in the device’s behavior.

Alongside the gain map, we also present a density plot to visualize how the clusters of Poincaré section points are distributed. Specifically, for each  $(P_{\text{pump}}, J_{c_2}) $, we calculate the standard deviation of the $V'_{\text{PS}}$ distribution, denoted as $\sigma_{V'_{\text{PS}}}$, see Fig.~\ref{Fig03}(b). A stable response corresponds to a very small value of $\sigma_{V'_{\text{PS}}}$ (i.e., $\sigma_{V'_{\text{PS}}}\ll0.01$~\cite{Note2}
, the yellow area), while a transition to chaos is marked by a sudden increase in $\sigma_{V'_{\text{PS}}}$ (i.e, $\sigma_{V'_{\text{PS}}}\gtrsim0.01$, the green-blue shaded area). The onset of the different regimes as $J_{c_2}$ changes is also clearly visible in this $\sigma_{V'_{\text{PS}}}$ map. \\

In summary, we have numerically studied the behavior of a JTWPA formed by junctions with a CPR that includes a second--harmonic contribution. The results show that, by adjusting the balance between the weights of the CPR harmonic components, it is possible to achieve a significant amplification, with gains up to $\sim 13\;\text{dB}$, even in the absence of dispersion engineering. Additionally, we have observed how non--sinusoidal contributions in the CPR can lead to a transition from stable, periodic behavior to chaotic dynamics, particularly for $J_{c_2} < 0$. These findings highlight the potential for optimizing JTWPA performance through the careful tuning of the harmonic components, offering an additional avenue for improving amplifier designs.

\begin{acknowledgments}
This work received support in part by PNRR MUR projects PE0000023-NQSTI (NbJTWPA and QSENS, Grant. No. H43C22000870001), in part by the Italian Institute of Nuclear Physics (INFN) through the DARTWARS and QUB--IT Projects, in part by the European Union’s H2020--MSCA under Grant 101027746, in part by the University of Salerno--Italy under Grants FRB21CAVAL, FRB22PAGAN, and FRB23BARON, and in part by PRIN 2022 PNRR Project QUESTIONs (Grant No. P2022KWFBH).
\end{acknowledgments}

\section*{Conflict of interest}
The authors have no conflicts to disclose.

\section*{Data Availability Statement}
The data that support the findings of this study are available from the corresponding author upon reasonable request.


%

\end{document}